\documentstyle[twocolumn,aps,prl,epsf]{revtex}
\begin{document}

\wideabs{
\title{The effect of disorder on the critical temperature of a
dilute hard sphere gas}

\author{M. C. Gordillo$^1$ and D. M. Ceperley$^2$}

\address{$^1$Departament de F\'{\i}sica i Enginyeria Nuclear, Campus
Nord B4-B5. Universitat Polit\`ecnica de Catalunya. E-08034
Barcelona, Spain}

\address{$^2$Department of Physics and NCSA, University of Illinois,
Urbana-Champaign, 61801}

\maketitle
\begin{abstract}
We have performed Path Integral Monte Carlo (PIMC) calculations to
determine the effect of quenched disorder on the superfluid
density of a dilute 3D hard sphere gas. The disorder was
introduced by locating set of hard cylinders randomly inside the
simulation cell. Our results indicate that the disorder leaves the
superfluid critical temperature basically unchanged. Comparison to
experiments of helium in Vycor is made.
\end{abstract}

\vspace*{3mm} \noindent PACS Numbers: 67.40.Hf, 67.40.-w}

Recently there have been several theoretical studies of the
superfluid transition (Bose condensation) of a system of
homogeneous hard spheres. For certain properties the model of hard
spheres can describe the Bose-Einstein condensates of alkali metal
atoms and liquid $^4$He. The ratio of the superfluid transition
temperature to that of an ideal gas $T_c/T_{c0}$ was determined
with numerical calculations \cite{peter} to be larger than unity
over a wide range of densities. The enhancement has been confirmed
by several other studies\cite{markus,baym} however the magnitude
of the enhancement as a function of density is still
controversial. It has also been proposed that T$_c$ can be
enhanced with respect to an ideal gas by the introduction of
quenched ({\it i.e.} static) disorder\cite{reppy1}. Experimental
results for $^4$He in Vycor show that, for some concentrations,
the enhancement is as large as 300 \%. Here, we study numerically
how a low density hard spheres gas is affected by random geometry.
Although there have been many calculations for disordered lattice
models \cite{li,moon}, to our knowledge, there are no other
calculations tackling this problem in the continuum.

Both helium gas and hard spheres system fall into the universality
class in which the critical exponents of the superfluid fraction,
$\varphi$, and the correlation length, $\nu$, satisfy $\varphi$ =
-$\nu$  = -0.67. In this class, the critical exponent of the
specific heat, $\alpha$ is negative and according to the Harris
criterium \cite{harris}, the presence of weak uncorrelated
disorder will not change the exponents. To examine both this
change in $\nu$ and the possible variation in T$_c$ with the
presence of quenched disorder, one can observe the behaviour of
helium when absorbed in a network of porous silica
\cite{reppy2,reppy3,reppy4,reppy5,reppy6,chan1}. This material is
produced with various porosities. For example, in aerogel the
volume of the voids in the system is around 90-95\%, in xerogel or
porous gold it is 60\%, and in Vycor the porosity is $\sim$ 30\%.
Experimentally, it is found that the introduction of disorder
decreases both the critical temperature and the fraction of the
helium atoms that decouple from the oscillator measurement of the
superfluid. In addition, the critical exponents can be changed:
for samples with large porosity, $\nu$ has a value greater than of
bulk helium case \cite{reppy4,reppy5}. However, in Vycor
\cite{reppy4} and porous gold \cite{chan2}, the exponents are the
same as in bulk $^4$He.

Since a system of homogeneous hard spheres belongs to the same
universality class as bulk liquid $^4$He, we tried to understand
the changes in the critical temperature and critical exponents in
silica gels by performing Path Integral Monte Carlo (PIMC)
calculations \cite{RMP} for a system of hard spheres with quenched
disorder. A simple model of disorder was constructed by placing,
at random, $N_c$ hard cylinders parallel to the edges of the cubic
periodic simulation cell. We used hard cylinders rather than a
more realistic model of Vycor or aerogel, because such a model is
much easier to treat computationally, in part because the absence
of an attractive interaction means that there will not be a ``dead
layer'' of $^4$He next to the substrate which would impose a
computational burden without participating in superflow. We chose
the positions of the cylinders by doing a simulation of a 2D hard
quantum hard disks. One snapshot was used to place the cylinders
along the x-axis, another independent snapshot for the y-axis and
a third for the z-axis. Thus there were always equal numbers of
cylinders along each axis.

The hard spheres are chosen to have diameter $\sigma$ appropriate
to $^4$He (2.2 \AA) so as to make comparison straightforward,
though we will also use dimensionless units. We demand that the
hard spheres remain a distance $d$ away from the center of each of
the cylinders.  The diameters of the cylinders were varied to
study the effect of their number and/or of the total excluded
volume. We considered two densities: $\rho \sigma^3=0.0107$ and
$1.07 \times 10^{-5}$.  Here, $\rho= N_s/L^3$, $N_s$ is the number
of hard spheres and $L$ the length of the simulation cell. A
summary of the conditions used is given in Table I. Each
simulation was repeated for five different realizations of the
disorder.

There are four dimensionless parameters characterizing this system
in the thermodynamic limit: the density of hard spheres
$\rho\sigma^3$, the fraction of excluded volume, an upper bound is
$\mu=\pi N_c (d/L)^2$ (an upper bound because if two cylinders are
close to each other, some of the excluded volume is counted
twice);  the ratio $d/\sigma$; and the ratio of the temperature to
that of free boson transition temperature at the same density
(T$_{c0}$=0.401 K at $\rho \sigma^3=0.0107$, not taking into
account excluded volume.)

First, let us examine how the precise geometry of the disorder
affects the superfluid fraction. Fig. 1 displays $\rho_s$/$\rho$,
for system II, as a function of T/T$_{c0}$.  The superfluid
density \cite{RMP} is computed by calculating the mean squared
winding of the paths around the periodic boundary conditions. The
crosses indicate the results for each disorder at a given
temperature, and the lines are the smoothed averages of the
superfluid fractions for the five cases. All the values of
$\rho_s$/$\rho$ are within a standard deviation of the average
values, indicating that the superfluid fraction does not depend
much on the particular geometry chosen and that one can speak of a
single T$_c$. Even though our systems small they seem to be
``self-averaging.'' The behaviour of system II is typical of the
other three geometries.

\begin{figure}
\epsfxsize=6cm  \epsfbox{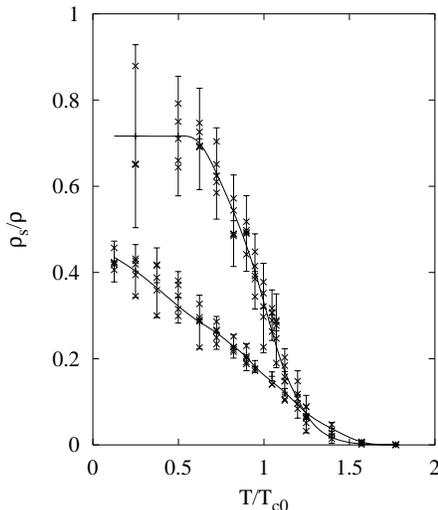} \caption{The upper curve shows the
superfluid fraction for system II with $N_c$=9 and $N_s$=118 The
symbols correspond to individual realizations, and the curve with
error bars to the disorder average. The error bars indicate one
standard deviation. The lower curve shows results for an ideal
gas.}
\end{figure}

The data in the lower part of the figure were obtained by
``switching off'' the interactions between the hard spheres and
keeping constant the rest of the simulation conditions, to gauge
the importance of the interaction on the results. We refer to this
as the ``disordered'' ideal gas. We did this for systems I,II and
III. It can be observed that when the hard core interaction is
turned off, the fraction of molecules belonging to the superfluid
at low temperatures decreases by a factor of two. In the
thermodynamic limit, the effect is even greater; the disorder
suppresses the superfluidity of the ideal gas. This can be easily
understood since, without interaction, there exists a lowest
energy state, localized in the cavity with the largest volume,
that will be occupied by all the bosons. Thus, when $N_s
\rightarrow \infty$, one will have BEC but not superfluidity. In
our simulations, we observed many exchanges, but in the absence of
interaction, the exchange loops remained in a single pocket. The
winding numbers were much less than in the interacting case and
were consistent with scaling to a non-superfluid state at all
temperatures.

Another  effect we observe in our simulation is the reduction of
superfluid density with increasing disorder. That can be seen in
Fig. 2 for the different systems with $N_s \sim $ 60. The
tortuosity, which we define as the fraction of atoms participating
in superflow at zero temperature, $\chi = \rho_s(T=0)/\rho$ given
in Table I, is found to be independent of $N_s$ and $0.51 \leq
\chi \leq 0.74$. These values are reasonably consistent with those
measured in Vycor $\chi = 0.76 \pm 0.01$\cite{reppy2}.  The
greater the excluded volume (system III vs. system II) for the
same type of cylinders, the lower $\chi$. For the same excluded
volume (systems I,II and IV), the superfluid fraction is greater
when the number of cylinders is smaller, if the diameter of the
hard spheres is comparable to that of the cylinders ($d \sim
\sigma$).

\begin{figure}
\epsfxsize=6cm  \epsfbox{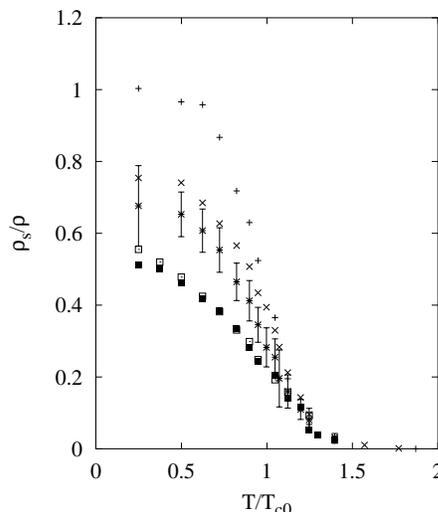} \caption{The superfluid density
for the pure system (pluses), system II (crosses), I (stars), IV
(open squares) and III (full squares). Here 54 $\le N_s \le$ 64.
The same ordering of the superfluid density occurred for other
values of $N_s$. The error bars are shown for one of the systems,
and are comparable in all other cases, except in the clean system,
which are a factor of two smaller.}
\end{figure}

To obtain an estimate of the superfluid transition temperature for
each system, we perform a finite size scaling analysis by fitting
$N_s^{1/3} \rho_s/\rho$ with \cite{pollock}:
\begin{equation}
A + B N_s^{(1/3 \nu)} (T - T_c) + C N_s^{-\Delta/(3\nu)}.
\end{equation}
This expression assumes the exponents for the superfluid density
and the correlation length are related by ($\varphi$ = -$\nu$).
The fit determines the critical temperature, T$_c$, and the
exponent for the correlation length, $\nu$. We assumed that
$\Delta=1/2$, but its precise value has very little effect on the
results. The range of reduced temperatures T$_1 \le$ T $\le$ T$_2$
( T$_1$ and T$_2$ given in Table II), included in the fit was
adjusted to ensure that the superfluid density was linear. The fit
for system I is given in Fig. 3. Shown in Table II are the
transition temperatures obtained with two different fits: one with
the $\nu$'s corresponding to the universality class of the hard
spheres (helium), and other choosing $\nu$ to minimize $\chi^2$.
Note that T$_c$ is independent of the value of $\nu$ chosen. The
introduction of the cylinders does not change T$_c$ much from an
homogeneous ideal gas at the same density. Recall that without
disorder T$_c$/T$_{c0}$= 1.057 $\pm$ 0.003 \cite{peter} (see
figure 4).

\begin{figure}
\epsfxsize=6cm  \epsfbox{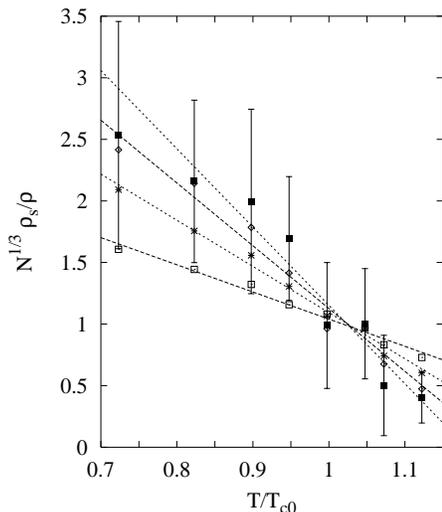} \caption{Scaled superfluid density
for system I. The symbols refer to $N_s$= 19, 54, 99 and 153. The
dotted lines are the fits and their intersection marks the
critical temperature for this system, in this case T$_c$/T$_{c0}$=
1.02 $\pm$ 0.05. The error bars are given for $N_s$=153, and are
at larger at least by a factor of two than in the other cases. }
\end{figure}

As the density of cylinders increases (system II vs. I) or their
diameter increases (system I vs. IV), the critical temperature
increases. This is most likely because as the available space for
the hard spheres decreases, their effective density increases. On
the other hand, for the same kind of cylinders (systems II and
III), T$_c$ does not change when the excluded volume does. Figure
(4) shows the comparison with the bulk system.  We see that within
error bars, disorder does not increase the transition temperature,
but on the other hand, it is not reduced either.

With respect to the universality class of the systems under
consideration, our conclusions are severely limited by the
statistical quality of the parameters obtained from the fit to Eq.
(1). For the most disordered system (III), the one with a excluded
volume of 0.33, we cannot draw any conclusion about the value of
$\nu$, so we cannot say if the universality class has changed.
Luckily, this fact is irrelevant for the critical temperatures. In
the other three systems, the situation is better. The error bars
of the critical exponent for the correlation length are small
enough to conclude that systems I and II belong to the same
universality class as pure helium (for which $\nu$ = 0.67). On the
other hand, in system IV $\nu$ = 1.32 $\pm$ 0.08. This value is
clearly different of 0.67, a result reinforced by the dramatic
decrease in $\chi^2$ when we go from $\nu = 0.67$ to the best fit
of $\nu= 1.32 $(see table II). This system might be in a different
universality class from pure helium. Such an increase of $\nu$ has
been experimentally observed in helium absorbed in aerogel and
xerogel \cite{reppy4,reppy5,chan1}.

We now compare these calculations to the measurements of the
transition temperature in Vycor. There are several important
differences. In Vycor the first helium atoms added to the system
are localized on the surface of the silica and do not contribute
to a superfluid response (the dead layer). The next helium atoms
form a surface layer around the silica strands which is locally 2D
but connects up at a longer length scale as a 3D network. The
pores are estimated to be in the range of 40 to 80 \AA.  For the
sample of ref. \cite{reppy1,reppy2} we estimate the excluded
volume, including that of the dead layer, is $\mu = 0.772$. This
is a much larger excluded volume factor and a much larger pore
size than in our simulations. The tortuosity of Vycor is estimated
to have a value of $\chi=0.76 \pm 0.01$ similar to that of our
systems.

\begin{figure}
\epsfxsize=8cm  \epsfbox{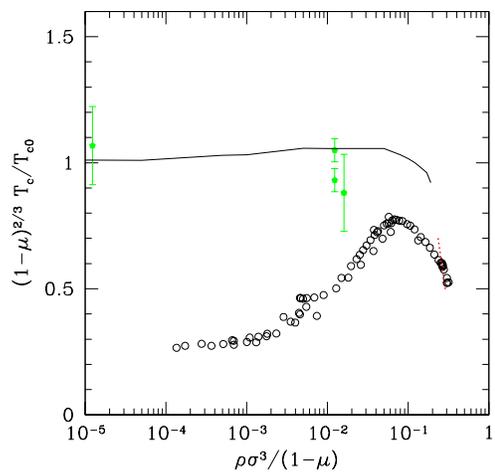} \caption{The superfluid transition
temperature versus the density. The solid line is the calculations
of ref. [1]. The $o$ are the data shown in fig. 3 of ref [4]. They
have been corrected for the free volume, tortuosity, and effective
mass. The solid symbols with error bars are the present
calculations. The dashed line is for bulk liquid $^4$He. }
\end{figure}

Figure 4 shows the comparison of the current results with the
experiment. To make this comparison, we corrected the densities
for the excluded volume in both sets of data. Plotted on the
horizontal axis is the density of helium in units of hard sphere
radii, defined as the number of free atoms (not including the dead
layer) divided by the free pore volume. The vertical axis is the
transition temperature divided by the ideal transition temperature
also taking into account the corrected density. The upper curve is
the estimate for pure hard spheres of ref \cite{peter}.  The
transition temperature in Vycor is reduced from the bulk hard
spheres value much more than in our calculation, by up to a factor
of three, most likely because of the density variation within a
pore and between pores. Assuming a pore size of 40\AA, for
densities such that $\rho \sigma^3/(1-\mu) \gtrsim 1.1 \times
10^{-3}$, one has more than one free atom/per pore, thus for these
densities, bose condensation can take place inside a single pore.
However, not until the pores are nearly full, is our model of
strictly repulsive disorder relevant for Vycor because of layering
within a pore. This effect, as well as the significant difference
in the ratio of pore size to hard sphere diameter, likely accounts
for the different reductions in the transition temperature at low
density. Also plotted on fig. (4) is the transition temperature of
bulk $^4$He versus density, showing essentially the same reduction
as for $^4$He in Vycor at the same density.

In conclusion we have performed PIMC simulation to determine the
superfluid density for hard sphere system outside of randomly
placed cylinders.  We have an overall reduction in the superfluid
density, however the taking into account the reduced free volume,
the transition temperature is reduced only slightly if at all.

This research was supported by NSF DMR-98-02373, NCSA for computer
facilities, and the Department of Physics at the University of
Illinois Urbana-Champaign. One of us (M.C.G.) thanks the Spanish
Ministry of Education and Culture for financial support. Comments
of J. Reppy, M. Holzmann and F. Lalo\"{e} and G. Baym are
appreciated.

\begin{table}
\begin{tabular}{lcccccc}
System &$\rho\sigma^3$&$\mu$ & $d/\sigma$&$\rho_s({\rm
T}=0)/\rho$& $N_{c}$ & $N_s$
\\ \hline

  I     &$1.07\times 10^{-2}$&0.128& 1 &0.68$\pm$0.10&       6       & 19 \\
        &      &    &    &  &                         12       & 54 \\
        &      &    &    &  &                    18       & 99 \\
        &      &    &    &  &                    24       & 153 \\ \hline

  II    &$1.07\times 10^{-2}$&0.128&1.5&$0.74\pm0.05$&       3       & 23 \\
        &      &    &    &    &                   6       & 64 \\
        &      &    &    &    &                   9       & 118 \\
        &      &    &    &    &                  12       & 182 \\ \hline

  III   &$1.07\times 10^{-2}$&0.33 &1.5&0.51$\pm$0.19&        9      & 29 \\
        &      &    &    &    &                   15      & 62 \\
        &      &    &    &    &                   21      & 102 \\
        &      &    &    &    &                   30      & 174 \\ \hline

  IV    &$1.07\times 10^{-5}$&0.128&15&  0.56$\pm$0.09&   3       & 23 \\
        &      &    &    &  &                     6       & 64 \\
        &      &    &    &   &                    9       & 118 \\
        &      &    &    &  &                    12       & 182 \\

\end{tabular}
\caption{Different cylinder arrangements and simulation conditions
used in this work.  Systems II and IV have the same randomness
(placement of the cylinders) }
\end{table}

\begin{table}
\begin{tabular}{lccccc}
System  & T$_1$/T$_{c0}$ & T$_2$/T$_{c0}$ & T$_c$/T$_{c0}$  &
$\nu$ & $\chi^2$
\\ \hline

  I     &  0.72 & 1.12 & 1.02 $\pm$ 0.05 &  0.67 (fixed)     & 0.8  \\
  I     &  0.72 & 1.12 & 1.02 $\pm$ 0.05 &  0.66 $\pm$ 0.05  & 0.8  \\ \hline

  II    &  0.72 & 1.20 & 1.15 $\pm$ 0.05 &  0.67 (fixed)     & 1.4  \\
  II    &  0.72 & 1.20 & 1.15 $\pm$ 0.05 &  0.71 $\pm$ 0.05  & 1.4  \\ \hline

  III   &  0.72 & 1.25 & 1.15 $\pm$ 0.20 &  0.67 (fixed)     & 0.7  \\
  III   &  0.72 & 1.25 & 1.12 $\pm$ 0.75 &  1.30 $\pm$ 0.5   & 0.4  \\ \hline

  IV    &  0.62 & 1.25 & 1.17 $\pm$ 0.05 &  0.67 (fixed)     & 14.5  \\
  IV    &  0.62 & 1.25 & 1.17 $\pm$ 0.17 &  1.32 $\pm$ 0.08  & 1.6  \\
\end{tabular}
\caption{ Transition temperature T$_c$ and exponent $\nu$ as
determined by the fit to Eq. (1). T$_1 \leq$ T $\leq $T$_2$ is the
fit range, $\chi^2$ is the reduced fit residual. The number of
data points in each fit was 32.}
\end{table}


\begin{references}


\bibitem{peter}
P. Gr\"{u}ter, D. M. Ceperley, and F. Lalo\"{e}, Phys. Rev. Lett.
{\bf 79}, 3549 (1997).


\bibitem{markus}
M. Holzmann and W. Krauth, Phys. Rev. Lett. {\bf 83}, 2687 (1999).


\bibitem{baym}
M. Holzmann, P. Gruter and F. Lalo\"{e}, Euro. Phys. J. B {\bf 10
}, 739 (1999); G. Baym, J. P. Blaziot, M. Holzmann, F. Lalo\"{e},
and D. Vautherin, Phys. Rev. Lett. {\bf 83}, 1703 (1999).


\bibitem{reppy1}
J.D. Reppy, B.C. Crooker, B. Hebral, A.D. Corwin, J. He and G.M. Zassenhaus,
Phys. Rev. Lett. {\bf 84}, 2060 (2000).

\bibitem{li}
Y.H. Li and S. Teitel
Phys. Rev. B {\bf 41}, 11388 (1990).

\bibitem{moon}
K. Moon and S.M. Girvin
Phys. Rev. Lett. {\bf 75}, 1328 (1995).


\bibitem{harris}
A. B. Harris, J. Phys. C {\bf 7}, 1671 (1974).


\bibitem{reppy2}
B.C. Crooker, B. Hebral, E.N. Smith, Y. Takano, and J.D. Reppy,
Phys. Rev. Lett. {\bf 51}, 666 (1983).


\bibitem{reppy3}
J.D. Reppy, Physica B+C {\bf 126}, 335 (1984).


\bibitem{reppy4}
M.H.W. Chan, K.I. Blum, S.Q. Murphy, G.K.S. Wong and J.D. Reppy,
Phys. Rev. Lett. {\bf 61}, 1950 (1988).


\bibitem{reppy5}
G.K.S. Wong, P.A. Crowell, H.A. Cho, and J.D. Reppy,
Phys. Rev. Lett. {\bf 65}, 2410 (1990).


\bibitem{reppy6}
P.A. Crowell, F.W. Van Keuls, and J.D. Reppy,
Phys. Rev. B {\bf 55}, 12620 (1997).


\bibitem{chan1}
J. Yoon, D. Sergatskov, J. Ma, N. Mulders, and M.H.W. Chan,
Phys. Rev. Lett. {\bf 80}, 1461 (1998).


\bibitem{chan2}
J. Yoon and  M.H.W. Chan, Phys. Rev. Lett. {\bf 78}, 4801 (1997).


\bibitem{RMP}
D. M. Ceperley, Rev. Mod. Phys. {\bf 67}, 279 (1995).


\bibitem{pollock}
E.L. Pollock and K.J. Runge,
Phys. Rev. B {\bf 46} 3535 (1992).


\end{references}
\end{document}